\documentclass{sig-alternate}
\usepackage[paper=letterpaper,width=7in,height=9.25in,centering]{geometry}
\newif\ifhepph
\hepphtrue
\newif\ifpublicationversion
\publicationversionfalse
\ifhepph\else\ifx\pdfoutput\undefined\else
\usepackage[%
  pdftex,%
  colorlinks=true,%
  hypertexnames=false,%
  \ifpublicationversion
    hyperfootnotes=false,%
    backref=false,%
    linkcolor=black,%
    citecolor=black,%
    anchorcolor=black,%
    filecolor=black,%
    menucolor=black,%
    urlcolor=black,%
  \else %
    backref=true,%
    citecolor=blue,%
  \fi %
  pdftitle={Automating Renormalization of Quantum Field Theories},%
  pdfdisplaydoctitle,%
  pdfauthor={A. D. Kennedy. Thomas Binoth, and Thomas Rippon},%
  pdfkeywords={Algorithms, Languages, Quantum Field Theory, Renormalization %
    Theory, Feynman Diagrams},%
  pdfstartview={FitH},%
  pdfpagemode={FullScreen}]{hyperref}
\usepackage{cite}
\fi\fi
\def\sharedaffiliation{%
\end{tabular}
\begin{tabular}{c}}
\def\union{\cup}		    
\def\Z{{\bb Z}}			    
\def\R{{\bb R}}			    
\def\C{{\bb C}}			    
\def\implies{\Rightarrow}	    
\def\SO{{\mathop{\rm SO}}}	    
\def\Spin{{\mathop{\rm Spin}}}	    
\def\L{{\cal L}}		    
\def\defn{\equiv}		    
\def\asymp{\sim}		    
\def\G{{\cal G}}		    
\def\H{{\cal H}}		    
\def\W{{\cal W}}		    
\def\SP{{\cal S}}		    
\input cyracc.def
\font\cyrxii=wncyr10 scaled900
\def\cyr{\cyrxii\cyracc}
\font\ninebb=msbm9
\font\sevenbb=msbm7
  \newfam\bbfam
  \def\bb{\fam\bbfam\ninebb}
  \textfont\bbfam=\ninebb
  \scriptfont\bbfam=\sevenbb
  \scriptscriptfont\bbfam=\sevenbb
\font\diagramx=diagram7
\font\diagramvii=diagram5
\font\diagramv=diagram5
\newfam\diagramfam

  \textfont\diagramfam=\diagramx
  \scriptfont\diagramfam=\diagramvii
  \scriptscriptfont\diagramfam=\diagramv
\def\diagchar#1{\count0="70 \advance\count0by\diagramfam %
  \multiply\count0by"100 \advance\count0by#1 \mathchar\the\count0}
\newcount\diagch
\def\diagop#1from#2to#3.{\diagch=#2 \diagchar\diagch%
  \loop\ifnum\diagch<#3 \advance\diagch by1 #1 \diagchar\diagch\repeat}
\begin{document}
\conferenceinfo{\it SNC'07,}{July 25--27, 2007, London, Ontario, Canada.} 
\CopyrightYear{2007}
\crdata{978-1-59593-744-5/07/0007} 
\ifpublicationversion
\else
  \permission{\copyright ACM, (2007). This is the author's version of the work.
    It is posted here by permission of ACM for your personal use.  Not for
    redistribution.}
\fi 
\title{Automating Renormalization of Quantum Field Theories}
\numberofauthors{3}
\author{
  \alignauthor A. D. Kennedy \\
  \email{adk@ph.ed.ac.uk}
  \alignauthor Thomas Binoth \\
  \email{Thomas.Binoth@ed.ac.uk}
  \alignauthor Tom Rippon \\
  \email{t.o.rippon@sms.ed.ac.uk}
  \sharedaffiliation
  \affaddr{School of Physics, University of Edinburgh} \\
  \affaddr{King's Buildings, Mayfield Road, Edinburgh, EH9 3JZ, Scotland}
}
%
\maketitle
\begin{abstract}
  We give an overview of state-of-the-art multi-loop Feynman diagram
  computations, and explain how we use symbolic manipulation to generate
  renormalized integrals that are then evaluated numerically.  We explain how
  we automate BPHZ renormalization using ``henges'' and ``sectors'', and give a
  brief description of the symbolic tensor and Dirac \(\gamma\)-matrix
  manipulation that is required.  We shall compare the use of general computer
  algebra systems such as {\tt Maple} with domain-specific languages such as
  {\tt FORM}, highlighting in particular memory management issues.
\end{abstract}

  [Physics]
\category{I.1.2,4}{Computing Methodologies}%
  {Symbolic and algebraic manipulation}[Algorithms, Applications]

\section{Introduction}
It has been said\footnote{R.~Jost} that ``in the thirties, under the
demoralizing influence of quantum-theoretic perturbation theory, the
mathematics required of a theoretical physicist was reduced to a rudimentary
knowledge of the Latin and Greek alphabets.'' Likewise, the ability to evaluate
multi-loop Feynman diagram can be reduced to an ability to do some simple
tensor algebra, some graphical manipulations to purge the calculations of
divergences, a knowledge of some basic properties of \(\Gamma\) functions, and
the ability to evaluate integrals numerically.  Since this is obviously
damaging to the delicate egos of theoretical physicists we advocate that the
entire procedure can --- and should --- be automated and handed over to
computers which do not have egos and, as is well-known, never make mistakes.

Precision measurements related to properties of elementary particles are
nowadays on a level which makes the inclusion of quantum corrections to
theoretical pre- and post-dictions mandatory.  As will be explained below these
quantum corrections are evaluated in the context of quantum field theoretic
perturbation theory, the loop expansion.  This allows for a systematic
evaluation of quantum corrections to --- in principle --- any order in the
coupling parameters of the theory under consideration.

\subsection{Loops and Legs: the State of the Art}
The complexity of perturbative calculations grows with the number of free
internal momenta which are integrated over (loops), and the number of external
particles of the process under consideration (legs).  Only a limited number of
observables have actually been calculated beyond the one-loop level.  For
one-loop scattering amplitudes on the other hand only one complete scattering
process with more that five external legs has been evaluated up to now: the
electroweak corrections to four fermion production in electron positron
collisions which is highly relevant for \(e^+e^-\)--collider phenomenology
\cite{Denner:2005es}.  For hadron colliders a lot of progress has been made
very recently concerning the evaluation of six-point amplitudes.  Different
methods have been designed to tackle this problem and benchmark amplitudes for
six-photon \cite{Nagy:2006xy, Binoth:2007ca, Ossola:2007bb} and six-gluon
scattering \cite{Britto:2006sj, Berger:2006ci, Ellis:2006ss, Bedford:2004nh,
Xiao:2006vt} are now available.  In the one-loop case so-called unitarity based
methods have played a prominent role in these developments (see
\cite{Bern:2007dw} for a review and references therein).  The unitarity
approach is not based on Feynman diagrams but rather on bootstrapping tree
level amplitudes.  Although it has a big potential its applicability to
amplitudes with internal and external masses needs further developments before
it may replace the Feynman diagrammatic approach.

It should be noted that the closed-form structure of one-loop amplitudes is
completely understood.  One can show that each one-loop Feynman integral is
expressible in terms of dilogarithms.  Even at two loops the situation is less
clear: only a limited number of four-point amplitudes, mainly for massless
particles, are known.  The corresponding Feynman diagrams are reduced first to
an independent set of basis functions, the so-called Master integrals, by using
integration by parts identities \cite{Chetyrkin:1981qh} and relations from
Lorentz invariance \cite{Gehrmann:1999as}.  The analytic knowledge of the
amplitude relies then on the successful evaluation of the Master integrals.
This is a formidable task and typically only successful if the number of scales
is very restricted.  For example a breakthrough in that direction was the
analytical evaluation based on Mellin-Barnes representations of the massless
planar \cite{Smirnov:2002kq, Heinrich:2004iq} and non-planar
\cite{Tausk:1999vh} four-point two loop integrals in 2000.  Although there has
been some progress for more complicated cases it has been very slow.
Unfortunately, even direct numerical evaluation is highly non-trivial as the
presence of infrared (IR) and ultraviolet (UV) divergences within the graphs
necessitates regularization; even in their absence integrable singularities due
to internal thresholds requires new numerical methods \cite{Binoth:2000ps,
Passarino:2001wv, Binoth:2003ak, Czakon:2005rk, Anastasiou:2007qb}.

\subsection{Beyond Two Loops}
Beyond the two loop level essentially only one-scale problems have been
evaluated so far.  A very remarkable result here is the recent analytical
evaluation of the three-loop splitting functions in QCD \cite{Vogt:2004mw,
Moch:2004pa}.  By mapping the splitting functions to Mellin moment integrals
one can generate a hierarchy of difference equations which have to be solved
recursively.  The project took many person-years and needed to accumulate a
database of integrals which occupied about 3.5 GBytes of disk space.  The
algebraic manipulations relied on the domain-specific language {\tt FORM}, or
more precisely on constant improvements of this code.  Indeed, the successful
completion of this important calculation relied on the fact that the author of
the algebraic manipulation code was a member of the collaboration.

The {\tt FORM} code was also the basis of the few known four-loop computations
in perturbative QCD.  All renormalization constants are known at this level
\cite{Chetyrkin:2004mf}.  An important milestone was in this respect the
evaluation of the four-loop \(\beta\)-function in QCD
\cite{vanRitbergen:1997va} in the so-called \(\overline{\mathop{\rm MS}}\)
renormalization scheme.  The problem was mapped to the evaluation of three loop
propagator functions which were computed with the {\tt FORM} package {\tt
MINCER}.  The project amounted to the evaluation of about 50,000 Feynman
diagrams and its confirmation by an independent group was an important issue
\cite{Czakon:2004bu}.  The latter mapped the problem to the evaluation of the
divergent part of some four-loop tadpole master integrals which were evaluated
by solving integration by parts identities using a dedicated {\tt C++} code.
To give a prominent example of a five-loop contribution to a very precisely
measured observable we should mention the numerical evaluation of the anomalous
magnetic moment of leptons by Kinoshita~et.~al.  Some dominant contributions
are known numerically now even at the five-loop level \cite{Kinoshita:2005sm}.
The parameter integrals corresponding to the Feynman diagrams are evaluated
with the Monte Carlo integration routine {\tt Vegas}.  This heroic effort will
presumably not be matched by an analytic computation in the foreseeable future.
Remarkably enough the three loop fully analytic result is available
\cite{Laporta:1996mq}.

\section{Quantum Field Theory}
When quantum field theories (QFTs) were developed in the 1930s they were
expressed in terms of field operators acting on Hilbert spaces and satisfying
various commutation and anticommutation relations.  Nowadays we prefer to think
of them in terms of functional (or path) integrals, which give us greater
geometrical insight (for example for the introduction of Fade'ev-Popov ghosts
for gauge-fixing non-abelian gauge theories) with just as high a level of
mathematical rigour (namely not very much).

\subsection{QFT as a Functional Integral}
The quantities we want to calculate are matrix elements of the scattering
(\(S\)) matrix, and these can be written as integrals over all possible field
configurations with a measure \(\exp(iS(\phi))\) where \(S = \int d^nx\,
\L(\phi, \partial\phi)\) is the action which is the space-time integral of the
Lagrangian density \(\L\) defining the theory.  We take the dimension of
space-time to be \(n\) rather than four as this is a convenient way of
``regulating'' the divergences that occur; when we have removed all the
divergences by some ``renormalization'' procedure that we shall discuss later
then we will take the limit that \(n\to4\) (or \(n\to6\) which is convenient
for the scalar \(\phi^3\) theory we are using as an example).  For a scalar
field theory, which for simplicity is all we shall consider here\footnote{The
reader should trust us that realistic theories with gauge fields, fermions,
ghosts and so forth just add some technical complications but do not
essentially alter the strategy and methods we propose.} the Lagrangian is \(\L
= \frac12\phi(\partial^2+m^2) \phi + \lambda\phi^3\).  A typical \(S\)-matrix
element is then \[\langle \phi(x)\phi(y)\,|\,S\,|\,\phi(z)\rangle = \frac1Z
\int d\phi\, \phi(x)\phi(y) \phi(z) e^{iS(\phi)},\] where the points \(x\) and
\(y\) are at time \(t=\infty\) and \(z\) is at \(t=-\infty\).  The integral is
over all fields \(\phi\): that is \(d\phi = \prod_x d\phi(x)\) where the
product is taken over all space-time points.  We shall not dwell on the precise
definition of such integrals but we will consider the situation where the
coupling constant \(\lambda\) is small and we can apply perturbation theory.
Formally one can consider that the weight factor \(\exp(iS)\) should be written
as \(\exp(iS/\hbar)\) in order to get the dimensions correct, and that as
Planck's constant \(\hbar\) is very small the integral will be dominated by
field values near to the minimum of the action which occurs at \(\phi = 0\).
In reality this is completely bogus, firstly because we are interested in
phenomena where the natural scale is set by working in units where \(\hbar = c
= 1\), and secondly because our \(\phi^3\) action is not bounded below as
\(\phi\to-\infty\).  To justify our argument we need to point out that we are
really expanding in the powers of \(\lambda\) rather than \(\hbar\), and that
for small perturbations about \(\phi=0\) the theory does not notice the
``non-perturbative'' instability of the vacuum.  We could try to be a little
less cavalier and consider a \(\phi^4\) interaction instead of or as well as
the \(\phi^3\) one, but it turns out that this theory is also sick, and so we
will keep things simple and just ignore these problems.

\subsection{Perturbation Theory}
The perturbative expansion we use is just an (unjustified) generalization of
the method of steepest descents to an infinite dimensional integral.  We add a
linear source term \(J\) to the action so that we can write \[\int d\phi\,
\phi(x) \phi(y)\phi(z) e^{iS(\phi)} = \frac{(-i)^3}Z \left.\frac{\partial^3
Z(J;\lambda)} {\partial J(x)\partial J(y) \partial J(z)}\right|_{J=0}\] where
\[Z(J;\lambda) \defn \int\!\!d\phi\,e^{iS(\phi) + iJ\phi} \asymp \exp\!
\left[i\int d^nx\, \lambda \frac{\partial^3}{\partial J(x)^3}\right] Z_0(J;0)\]
where we have written \(J\phi\) for \(\int d^nx\, J(x)\phi(x)\), and \(Z_0(J)
\defn \int d\phi\,e^{i\left(\frac12 \phi K\phi + J\phi\right)} \propto
e^{-\frac i2 J\Delta J}\), where we have introduced the kernel \(K = \partial^2
+ m^2\) and its Green's function \(\Delta \defn K^{-1}\) with appropriate
boundary conditions.  We obtain the required asymptotic expansion by expanding
\(\exp\left(i\lambda\,\partial^3/\partial J^3\right)\) as a Taylor series, and
then each resulting term can be drawn as a Feynman diagram with the rules that
there is a factor of \(i\lambda\) associated with each vertex whose position is
integrated over all space-time locations, and a propagator \(i\Delta\) with
each edge.  If we Fourier transform to momentum space then we obtain the
equivalent diagrams but with the rules that momentum is conserved at each
vertex and there is an integral over the momentum flowing round each loop, and
the propagators become \(\Delta(k) = (k^2-m^2+i0_+)^{-1}\).

\section[Divergences and Renormalization]{Divergences and\\ Renormalization}
The infamous UV divergence disease of QFT is now immediately apparent, it
manifests itself even if we work in Euclidean rather than Minkowski space.
Even the simplest one-loop diagram corresponding to the Euclidean integral
\[I(p) \defn \int \frac{d^nk}{[k^2+m^2][(k+p)^2+m^2]} < c\int_0^\infty
\frac{dk\,\|k\|^{n-1}}{\max(\|k\|, m)^4}\] for some constant \(c\), and this
clearly diverges in \(n\geq4\) dimensions.

What saves the day is that the divergence is local (a Dirac \(\delta\)-function
or its derivatives in space-time, or equivalently a polynomial in the external
momentum \(p\) in momentum space).  We may see this easily by differentiating
the Feynman integral with respect to \(p\): \[\frac{\partial I(p)}{\partial
p_\mu} = -\int \frac{d^nk\,2(k+p)_\mu}{[k^2+m^2] [(k+p)^2+m^2]^2}.\] For large
\(\|k\|\) the integrand now behaves as \(\|k\|^{-5}\) rather than
\(\|k\|^{-4}\) as before; differentiating \(d = n-3\) times suffices to render
the integral convergent.  We may therefore express the Feynman integral as
\[I(p) = T^d I(p) + \int_0^p d^np_1\cdots \int_0^{p_{d-1}} d^np_d\, \partial^d
I(p_d)\] by iterating the fundamental theorem of calculus \(d\) times
(otherwise known as Taylor's theorem) where \(T^d I(p)\) is a polynomial of
degree \(d\) with divergent coefficients.  We have simplified the notation by
writing \(\partial\) for \(\partial/\partial p_\mu\), suppressing all the
tensor indices (such as \(\mu\)), and lumping together all the external momenta
into one \(p\).

We can remove the offending divergent polynomial by add\-ing it to the action
as a new term giving rise to a vertex that is formally of one-loop order.  The
number of loops in a Feynman diagram is formally equal to the corresponding
power of \(\hbar\) in the perturbation expansion, so we may choose to imagine
that all the couplings in the action, \(\lambda\), \(m\), and the coefficient
\(Z\) of the kinetic term \(\phi\partial^2\phi\) that we forgot to include
originally, may be expanded in powers of \(\hbar\) where all the terms but the
first just serve to remove the unwanted divergences.  Clearly there are two
requirements that must be satisfied for this to work: (i)~all the divergences
must be of the form of terms that occur in the action, and (ii)~all the
divergences must add up in just the right way to correspond to such
counterterms.  We usually rephrase (i) by saying that we must include all
monomials in the action that are allowed (i.e., which do not violate physically
necessary conditions such as locality and unitarity), as if we do not we will
have to come back later and put them in so as to cancel the divergences that
arise.  Of course, we can only allow there to be a finite number of parameters
(and thus monomials) in the action if we are to have a theory with any
predictive power, but happily we can get away with including only monomials of
dimension less than \(n\), as this suffices to cancel all possible divergences
by a simple power-counting argument~(q.v.,~\S\ref{sec:PowerCounting}).

\subsection{Regulators}
Of course this is all rather messy and embarrassing, so we reformulate the
procedure in the following language: in order to define a QFT we need to
introduce a regulator of some kind which makes our manipulations mathematically
well-defined.  We then renormalize the theory by adjusting the ``bare''
coefficients in the action to absorb all the would-be divergences, and finally
we take the regulator away to obtain a well-defined finite theory as the limit
of a renormalized regularized functional integral.  There are many choices of
regulator, working in \(n\) dimensions (``dimensional regularization''
\cite{thooft:1972fi, thooft:1973a, bollini:1972ui, cicuta:1972a}) is just one
rather convenient possibility, and they are all supposed to give the same
answer.  What turns out to be crucial is that the regulator preserves as many
of the symmetries of the original theory as possible, as we can then exclude
counterterms that do not also have these symmetries.

\subsection[Renormalization Conditions and the Renormalization
  Group]{Renormalization Conditions and\\ the Renormalization Group}
In order to fix the finite parts of the counterterms in the action we need to
specify a set of renormalization conditions, i.e., a set of experimentally
determined values for some processes that can be solved for the parameters in
the action. This is no different from what happens in classical theories,
except that the parameters themselves have a less obvious physical meaning.
Indeed, we can choose the renormalization conditions in many different ways;
for example we can specify them as a function of some energy scale \(\mu\), and
the fact that they all fix the same physical theory tells us the physical
quantities must not depend upon \(\mu\). The invariance under this group of
reparameterizations is known as renormalization group invariance, and is very
useful because it is not in general respected by the perturbative expansion to
any given order.

The parameters in the action that are just constants on the classical level
thus become in general scale dependent if radiative corrections are included.
In the context of Quantum Chromodynamics (QCD), our theory of the strong
interactions, this leads to asymptotic freedom, the logarithmic decrease of the
coupling strength with increasing interaction energy.  For QCD the perturbative
approach is limited to the high energy regime where the gauge coupling
parameter, \(\alpha_s\) is sufficiently small.

In the case of gauge theories (which are the theories we really use to describe
nature) we must choose renormalization conditions that preserve the gauge
symmetries or the whole theory falls apart.  On the other hand there are
certain symmetries, such as some forms of chiral symmetry (a peculiar symmetry
that occurs because our theories involve fields that transform as spinor
representations of the covering group \(\Spin(3,1)\) of the Lorentz group
\(\SO(3,1)\)) and scale invariance symmetry cannot be maintained by any
regulator and thus are not symmetries of the quantum theory even though they
look like perfectly good symmetries of the underlying classical action.  There
are physical implications of these so-called anomalies, such as the decay of a
pion into two photons \(\pi\to\gamma\gamma\), and this is evidence that
renormalization is necessary and not just an ugly contrivance that we could
avoid by being more clever.

What will concern us for the rest of this paper is a more detailed
investigation of~(ii), namely showing that all the divergences are local and
appear in the right way to be cancelled by counterterms in the action.  For
instance, if we go beyond one-loop Feynman diagrams are all the divergences
still local? The trouble is that multi-loop Feynman diagrams are very
complicated integrals, and it is hard to see what is going on without getting
lost.  Nevertheless, with a little effort one can see that a typical two loop
diagram \(\diagchar{'00}\) has non-local divergences; however, before
abandoning all hope we notice that the one-loop counterterms we introduced to
cancel the one-loop divergences have to appear in loop diagrams
themselves.  Their divergent coefficients thus multiply the non-local finite
parts of the corresponding Feynman integrals also leading to non-local
divergences.  What we must show is that these one loop countergraphs, which are
formally of two loop order (\(O(\hbar^2)\)) exactly cancel the non-local part
of our two loop graphs, leaving only a local divergence that can be absorbed by
adding local two loop counterterms to the action.

\subsection[Henges and the R and R-bar Operations]{Henges and the
  {\large\(\mathbf R\)} and {\large\(\mathbf{\bar R}\)} Operations}
  \label{sec:R} 
This is most easily done in two stages.  The first is we define a procedure
that removes all divergences by local Taylor series subtractions, and then we
show that the subtractions thus made add up to give local counterterms.  We
will discuss the first stage here, as it is what needs to be implemented to
automate the renormalization process; the second is a purely combinatorial
proof which the interested reader can find in the literature~\cite{kennedy82b,
anikin:1976a, collins84a}.

The two-loop diagram \(\diagchar{'00}\) illustrates the difficulty we face.
Clearly when all the loop momenta get large simultaneously the diagram diverges
as \(\|k\|^{2n-10}\), and this overall divergence is local.  However, what
happens when one of the two three-line loop momenta gets large while the other
stays small? How do we disentangling these overlapping subdivergences?  Our
approach is a systematic decomposition of the space of all loop momenta based
on the structure of the graph itself.

We need only consider graphs which are one-particle irreducible (1PI), that is
ones which remain connected when any line is cut.  For any line \(\ell\) in a
1PI Feynman diagram \(\G\) we can decompose the graph uniquely into a single
loop containing \(\ell\) stringing together a set of 1PI subgraphs that we will
call a Henge \(\H(\H,\ell)\) \cite{kennedy83a, kennedy96d}.  We will discuss
efficient ways of representing and computing Henges in~\S\ref{sec:Algorithms}.

Clearly at any point the space of all loop momenta some line must carry the
smallest (Euclidean) momentum.\footnote{Up to a set of measure zero.} Hence we
may decompose an arbitrary Feynman diagram \(\G\) into a sum of contributions
each from a region where a particular line \(\ell\in\G\) carries the smallest
momentum, and in each such region we can use our Henge decomposition to write
the Feynman integral as \[I_\lambda(\G) = \sum_{\ell\in\G} \int_\lambda^\infty
d^nk_\ell\, i_{k_\ell}\Bigl(\G/\H(\G,\ell)\Bigr) \prod_{\Theta\in\H(\G,\ell)}
I_{k_\ell}(\Theta),\] where \(I_\lambda(\Theta)\) is the Feynman integral for
the graph \(\Theta\) restricted to the region of momentum space where all the
lines carry momenta of magnitude greater than \(\lambda\), and
\(i_\lambda(\G/\H)\) is the product of all the lines in the single loop
\(\G/\H\) obtained by shrinking all the graphs in \(\H\) to points, again
restricted to have momentum of magnitude greater than \(\lambda\).  The ``small
momentum cutoff'' \(\lambda\) serves as a convenient parameter for our
recursive definition.

As an example let us consider the overlapping two-loop diagram we considered
before.  Its concomitant Feynman integral is \[I(p) = \int d^nk\, d^nk'\,
\Delta(k) \Delta(k+p) \Delta(k-k') \Delta(k'+p) \Delta(k').\] The Henges that
arise are \(\diagchar{'01}, \diagchar{'02}, \diagchar{'03}\), where the heavy
lines show the Henges corresponding to each of the light lines.  For example,
the contribution from the region where the light line in the first diagram is
smallest is \[I_\lambda\left(\diagchar{'14}\right) = \int_\lambda^\infty
d^nk''\, \Delta_\lambda(k'')\; I_{k''}\left(\diagchar{'01}\right)\] where
\(\Delta_\lambda(k) \defn \Delta(k) \theta\bigl(\|k\| - \lambda\bigr)\) and the
inner integral (corresponding to the blob in the diagram on the left and the
heavy loop in that on the right) is \[\int_{\|k'' \|}^\infty d^nk\,
\Delta_{k''}(k) \Delta_{k''}(k+p) \Delta_{k''}(k + k'' + p) \Delta_{k''}(k +
k'').\] Observe how the graphical structure dictated the change of variable
\(k'' = k-k'\) to the loop momentum of the shrunken graph.

With this decomposition it is obvious that the divergences can be put into two
classes, overall divergences that occur when all the momenta simultaneously get
large, and subdivergences that are isolated to the subgraphs occurring in the
Henges.  Let us introduce an operator \(R\) that removes all the divergences
from a given Feynman diagram: it first removes all the subdivergences (an
operation that is called \(\bar R\)) by recursively applying \(R\) to the
elements of the Henges \[\bar RI_\lambda(\G) \defn \sum_{\ell\in\G}
\int_\lambda^\infty d^nk_\ell\, i_{k_\ell}\Bigl(\G/\H(\G,\ell)\Bigr)
\prod_{\Theta\in\H(\G,\ell)} RI_{k_\ell}(\Theta),\] and then it removes the
overall divergence (if necessary) by subtracting the leading terms of the
Taylor series expansion in the external momenta as described before for the
one-loop case, \(RI_\lambda(\G) \defn I_\lambda(\G) - T^{\deg(\G)}\bar
RI_0(\G)\).  The number of terms removed by the Taylor subtraction operation
\(T\) is fixed by simple power counting rules (q.v.,
\S\ref{sec:PowerCounting}).  Note the subtle but vital fact that the
subtraction term has its small momentum cutoff set to zero.

\subsection{BPH and Zimmermann's Forests}
In order for this to be a valid renormalization procedure we need to prove two
things: first that all the overall divergences are local --- that is polynomial
in the external momenta --- so that they get eaten by the Taylor series
subtraction, and second that all the subtractions that are made add up in such
a way as to correspond to counterterms in the action.  The former condition is
equivalent to showing that the derivatives \(\partial^{\deg(\G)+1} \bar
RI_\lambda(\G)\) with respect to the external momenta are finite, and follows
by induction from (a)~the inductive assumption that \(|I_\lambda(\G)| < c\cdot
\max(\lambda, m)^{\deg(\G)}\) for some constant \(c\), (b)~the requirement that
differentiation lowers the power-counting degree, \(\deg(\partial\G) \leq
\deg(\G) - 1\), and (c)~the fact that differentiation commutes with the \(R\)
operation, \([\partial,R]=0\).  To prove~(c), as well as the condition~(ii)
above, we rewrite the \(\bar R\) operation in purely graphical form (as it was
originally defined by Bogoliubov and Parasiuk\footnote{Their original proof was
corrected later by Hepp, hence it is known as the BPH theorem
\cite{bogoliubov:1957a, parasiuk:1960a, hepp:1966a, anikin:1973a, kennedy82b,
kennedy96d}}), namely \[\bar RI(\G) = \sum_{\SP\in\bar\W(\G)} I(\G/\SP)*
\prod_{\Gamma\in\SP} \left(-T^{\deg(\Gamma)} \bar RI(\Gamma)\right).\] Here
\(\SP\) is a spinney, that is a covering of \(\G\) by a set of 1PI subgraphs
(possibly including single vertices), and the proper wood \(\bar\W\) is the set
of all such spinneys (excluding the one consisting of just \(\G\) itself).  The
notation \(I(G/\Gamma)*f(\Gamma)\) means that the 1PI subgraph \(\Gamma\) in
\(\G\) is to be shrunk to a point and replaced with the value \(f(\Gamma)\).

Proving the equivalence of this definition of \(R\) to the previous one for
\(\lambda=0\) is left as an exercise for the dedicated reader who need just
show that in the recursive Henge decomposition the sequence of subtractions
made always correspond to some spinney, and that for each spinney \(\SP\) the
corresponding subtractions occur in the Henge decomposition for every ordering
of the line momenta in \(\G/\SP\).

One can expand out the recursion even more and obtain an explicit formula for
all the subtractions that are made in a graph in terms of nested
non-overlapping Taylor subtractions; the necessary graphical apparatus for this
was given by Zimmermann \cite{zimmermann:1971a} in his forest formula.
Zimmermann introduced his forest formula because it explicitly defines a
convergent integral corresponding to a Feynman diagram without the need for any
intermediate regularization scheme.  This property is the reason for our
practical use of the \(R\) operation: we use it to generate the integrand of a
manifestly convergent integral, although in practice this is more easily
implemented using Henge recursion.  From a formal point of view the utility of
Zimmermann's approach is limited by the fact that although it defines finite
integrals these are not directly related to the original functional integral in
a straightforward way, and therefore the various identities between Feynman
integrals that follow because of this that are needed to guarantee desirable
properties such as gauge invariance (Ward or BRS identities) or unitarity
(cutting relations) are no longer manifest.  Indeed the former are still untrue
in the case of anomalies despite the fact that no regulator is introduced.

The fact that Bogoliubov's definition is purely graphical makes it obvious that
\([\partial,R]=0\), and it is also possible to give a combinatorial argument
that shows that all the subtractions made, summed over all Feynman graphs,
formally corresponds to adding \(-T\bar Re^{i\lambda\phi^3}\) to the
action~\cite{kennedy82b, anikin:1976a}.

\subsection{Power Counting} \label{sec:PowerCounting}
The BPH theorem tells us how to remove all the divergences from a Feynman
diagram by local subtractions, but it does not guarantee that the number of
counterterms required is finite.  If it is not then the resulting theory has
little if any predictive power, as an infinite number of renormalization
conditions will be needed to fix the finite parts of all these new vertices in
the theory.  The class of theories that can be renormalized with a finite
number of counterterms, called renormalizable theories, is therefore of central
interest.  We can give a simple power counting rule that tells us when a theory
is renormalizable (but not the converse).

The \(\deg\) function introduced in \S\ref{sec:R} must be chosen such that
\(|\Delta(k)| \leq c\cdot\max(k, m)^{\deg(\Delta)}\) for all vertices,
propagators (lines), and their derivatives in the Feynman rules associating
integrals with graphs; this is the starting point for the inductive proof that
\(|I_\lambda(\G)| \leq c\cdot\max(\lambda, m)^{\deg(\G)}\) that is used to
establish the BPH theorem in our approach.  For our working example of
\(\phi^3\) theory in \(n\) dimensions the degree of each propagator is \(d=-2\)
and the vertices have degree \(d'=0\).  The degree of a graph with \(I\)
internal lines, \(V\) vertices, and \(L\) loops is thus \(\deg = Id + Vd' +
Ln\), recalling that there is an \(n\)-dimensional momentum integration
associated with each loop.  Now, any connected graph with \(V\) vertices has a
spanning tree containing exactly \(V-1\) edges, and the remaining \(I-V+1\)
edges each give rise to an independent loop; furthermore each internal line has
two ends and, in our theory, each vertex connects three lines, so
``conservation of line ends'' implies that \(3V = 2I+E\) where \(E\) is the
number of external lines.  Eliminating \(L\) and \(I\) from these equations we
find that \(\deg = n - E\frac12(d+n) + V\frac12(2d'+3d+n) \defn n - E\dim(\phi)
+ V\dim(V)\) where the dimension of the field \(\dim(\phi) = \frac12(n-2)\) and
of the vertex is \(\dim(V) = \frac12(n-6)\).

These dimensions can be read directly from the Lagrangian describing the theory
by some simple rules.  We see that if \(\dim(V) > 0\) then we can find
arbitrarily overall divergent graphs just by increasing the number of vertices
sufficiently, whereas if \(\dim(V) < 0\) then only a finite number of graphs
can have overall divergences (such theories are called superrenormalizable).
The case \(\dim(V) = 0\) (renormalizable) is particularly interesting as there
can be an infinitude of overall divergent graphs, but as their degree is
bounded by \(E\dim(\phi)\) (provided this is positive) only a finite number of
Green's functions are overall divergent.  In four dimensions \(n=4\) we have
\(\dim(\phi)=1\) and \(\dim(V)=-1\) for \(\phi^3\) theory, so it is
superrenormalizable, whereas in six dimensions \(\dim(\phi)=2\) and \(\dim(V)
=0\) it is renormalizable.

\section[Numerical versus Analytic Evaluation of Integrals]{Numerical versus
  Analytic\\ Evaluation of Integrals}
Heretofore people have usually calculated Feynman integrals by using some
regulator to make all the manipulations well-defined and then arranged the
resulting expressions to cancel the would-be divergences between graphs and
countergraphs, leaving answers with a finite limit as the regulator is removed.
This procedure is effective if all the integrals can be evaluated in closed
form,\footnote{In fact, what is really needed is that the would-be divergent
parts can be evaluated in closed form.  The usual folklore is that the work
required to evaluate the divergent parts of diagrams with \(\ell+1\) loops is
about the same as that to evaluate the finite parts with \(\ell\) loops.} but
it is very difficult to completely automate the procedure as the closed-form
evaluation of the integrals is a new challenge at each order of the loop
expansion.

We wish to use the \(R\) operation to generate manifestly finite Feynman
integrals, which we can then evaluate numerically.  The advantages of this
approach are obvious, but there is a price to pay.  First, we need to evaluate
the albeit finite integrals for every value of the external momenta and
parameters (renormalization conditions) separately.  Second, the integrals are
over a unit cube in \(I-1\) dimensions (q.v., \S\ref{sec:Parameters}) where
\(I\) is the number of edges in the graph, and thus we have to resort to Monte
Carlo methods in all but the most trivial cases, so we are only able to achieve
limited precision.  Third, in most cases where perturbation theory is
applicable we are expanding in a small parameter (the fine structure constant
\(\alpha \approx\frac1{137}\) in QED for example), so there is no point in
computing the \(\ell+1\) loop contributions unless the errors at \(\ell\) loops
are sufficiently small.  Our window of opportunity is to compute one loop
higher order that the current best closed-form solutions.

\section{Graphical Algorithms} \label{sec:Algorithms}
We now consider how we can implement the \(R\) operation in practice in our
symbolic--numeric scheme.  It is much easier to implement the more recursive
Henge approach than Bogoliubov's definition or Zimmermann's explicit forest
formula, as the complexity of the problem is then naturally handled by a
correspondingly recursive program.  To do this we need some reasonably
efficient graph manipulation algorithms, as the text-book methods using
adjacency matrices and suchlike are slow in practice.  Fortunately such
algorithms are well-known, such as the Galler--Fisher \cite{galler:1964a,
knuth:1997a} algorithm for finding equivalence classes given pairwise
equivalence relations, which may be equivalently viewed as a means of splitting
a graph into its connected components.

The basic idea is to represent a graph as a set of edges each represented as a
pair of vertices, and to label the vertices with small integers; for instance
the graph \(\diagchar{'00}\) could be represented as \(\{\langle12\rangle,
\langle23\rangle, \langle34\rangle, \langle41\rangle, \langle24\rangle\}\).  We
create an array \(F\) indexed by vertices and initialised to some special value
(such as zero), and then run once through the edges \(\langle ij\rangle\)
computing the ancestor of each of its vertices, where the ancestor \(A(i)\) is
\(i\) if \(F[i] = 0\) or \(A(F[i])\) otherwise.  If the ancestors are unequal
\(A(i)\neq A(j)\) then we set \(F[A(i)] \gets A(j)\).  When we have exhausted
the supply of edges the ancestor of any vertex labels the connected component
that the vertex belongs to.

This algorithm is pleasantly easy to extend to carry out other graphical tasks
efficiently.  To find all the irreducible edges in a connected graph (those
that may be cut without disconnecting it) we just keep track of the route from
a vertex to its ancestor by introducing an \(\Z_3\) chain\footnote{A \(\Z_3\)
chain is a sum of edges with coefficients~\(\pm1\) or~\(0\).}-valued array
\(P\) indexed by vertices that contains the chain of edges connecting \(i\) to
\(F[i]\), and defining the route \(R(i)\) from \(i\) to \(A(i)\) to be zero if
\(i=A(i)\) or the chain \(R(i) = P[i] \oplus R(F[i])\) otherwise.  Upon
examining the edge \(\langle ij\rangle\) we note that the chain \(C_{ij} \defn
R(j)\oplus\langle ij\rangle\ominus R(i)\) connects \(A(i)\) to \(A(j)\), so if
\(A(i)\neq A(j)\) as well as updating \(F[i]\) as above we also set \(P[i]\gets
C_{ij}\).  If, on the other hand, \(A(i)=A(j)\) then we have discovered that
\(C_{ij}\) is a closed loop, and all the edges occurring in this loop are
marked as irreducible.  All the edges not so marked by the end of the loop over
edges are reducible.

One use of this algorithm is to route momenta through a graph.  We multiply
each loop by a symbol for the corresponding loop momentum, and for each vertex
\(i\) at which external momentum \(p\) enters the graph we multiply the route
\(R(i)\) by \(p\).  The coefficient of any edge in the sum of all these
\(p\)-chains\footnote{A \(p\)-chain consists of sums of edges with coefficients
being \(\pm\) symbolic names for momenta.} is the momentum flowing through that
edge.

Indeed, it is but a small step further to devise an efficient Henge-finding
algorithm.  Start with a 1PI graph \(\G\) and remove any edge~\(\ell\); as
\(\G\) is just a set of edges this graph is just the set of edges \(\G -
\{\ell\}\).  Now apply the previous algorithm to this graph, partitioning the
lines into two sets \(\G = I\union R\) of irreducible lines and reducible
lines, and apply the first algorithm to partition \(I\) into disconnected
pieces \(I = \{\theta\}\).  The required Henge \(\H(\G,\ell)\) is just this
set, and the single loop \(\G/\H(\G,\ell) = R\union \{\ell\}\).

Using these algorithms we have implemented the subtractions specified by the
\(R\) operation on an arbitrary diagram in \(\phi^3\) theory, and the
generalization to more interesting theories with more complicated actions is
both straightforward and in progress.  It is perhaps worth noting for the
benefit of the dedicated reader who has thought about the equivalence of the
Henge and Bogoliubov definitions of \(R\) that too na{\"\i}ve an implementation
of the former will produce the same spinney multiple times, corresponding to
different orderings of lines in \(\G-\SP\).

\section{Feynman Parameters} \label{sec:Parameters}
We have described in some detail the means of obtaining a finite Feynman
integral, but we still have to evaluate it.  There are various ways of doing
this: we could for example directly evaluate the loop integrals in momentum
space, leading to an \(nL\)-fold real integral.  For some regulators, such as
the lattice, where we discretize the functional integral by approximating
space-time by a hypercubic grid, this is essentially the only way to proceed.
In practice, however, we usual carry out perturbative QFT calculations using
dimensional regularization in which case all of the propagators are essentially
inverse quadratic forms in the momenta, as we used above.  In this case there
is a convenient \(\Gamma\) function identity that allows us to write a product
of such propagators as a power of a single quadratic form
\[\frac{\Gamma(\alpha) \Gamma(\beta)}{A^\alpha B^\beta} = \Gamma(\alpha+\beta)
\int_0^\infty \frac{dt\,t^{\beta-1}}{(A+Bt)^{\alpha+\beta}}.\] This can be
iterated to establish the formula
\begin{equation}
  \prod_{i=1}^N \frac1{Q_i}
    = (N-1)! \int_0^1 dx_1 \cdots \int_0^1 dx_N
      \frac{\delta\left(1 - \sum_{k=1}^N x_k\right)}{\left[\sum_{j=1}^N
        x_jQ_j\right]^N},
  \label{eq:FeynmanParameters}
\end{equation}
the quantities \(x_i\) being known as
Feynman parameters.

Introducing Feynman parameters and interchanging the order of the momentum and
parameter integrals (which is valid in the presence of a regulator) we can
combine all \(nL=2\omega\) loop momenta into a single vector and carry out the
momentum integration using another \(\Gamma\) function identity, \[\int
\frac{d^{2 \omega} k}{\bigl[k^2 + F(p,m)\bigr]^\alpha} =\pi^\omega
\frac{\Gamma(\alpha -\omega)}{\Gamma( \alpha)} F(p,m)^{\omega-\alpha}\] and
related formul\ae\ obtained from this by differenting with respect to the
external momentum~\(p\).  We are then left with \(I\) parameter integrals to
evaluate, in our case numerically.

\subsection{Sectors}
It is interesting to see how the Henge decomposition is reflected in Feynman
parameter space.  To this end we define a sector of Feynman parameter space as
a region where the parameters are totally ordered, for example one such sector
is \(x_1>x_2>\cdots>x_N\).  If we restrict the integral in
equation~(\ref{eq:FeynmanParameters}) to this sector we find
\begin{eqnarray*}
  && \int_0^1 dx_1 \int_0^{x_1} dx_2 \cdots \int_0^{x_{N-1}}\!\!dx_N
    \frac{\delta\left(1-\sum_{k=1}^N x_k\right)}
	 {\left[\sum_{j=1}^N x_jQ_j\right]^N} \\
  && \hskip4cm = \frac1{(N-1)!} \prod_{i=1}^N\frac1{\sum_{j=1}^iQ_j},
\end{eqnarray*}
so for example the contribution to the product \(\frac1{TUV}\) represented
using Feynman parameters \(t\), \(u\), and \(v\) from the sector \(t>u>v\) is
\(\frac1{T(T+U)(T+U+V)}\).  This means that the quadratic form denominator
\(T\) corresponding to parameter \(t\) is guaranteed to be smaller than all the
other denominators, which means that it corresponds to a line carrying the
smallest momentum in the Henge decomposition.

We note in passing that especially in the context of overlapping IR divergences
the the iterated application of sector decomposition has proved to be a
powerful way to deal numerically not only with multi-loop Feynman diagrams
\cite{Binoth:2000ps, Binoth:2003ak} but also high-dimensional phase-space
integrals~\cite{Binoth:2004jv}.

\section[Tensor and gamma-Matrix Manipulation]{Tensor and
  {\large\(\mathbf\gamma\)}-Matrix\\ Manipulation}
In this pedagogical presentation we have illustrated our calculations using
only scalar fields; in general we need to introduce vector fields for gauge
bosons and spinor fields for fermions, these lead to the additional
complication of having tensor expression in the Feynman integrand numerators.
Actually our symbolic manipulation code already has to deal with such
numerators, as they occur when we Taylor subtract even for scalar theories.
What we end up with is some complicated \(\SO(4)\) tensor expressions which
need to be simplified into some canonical form; these expressions can
themselves be viewed as graphs, where the tensors are the vertices and the
edges are pairs of contracted indices, and we use the algorithms of
section~\S\ref{sec:Algorithms} to carry out this simplification.

For spinors, which strictly speaking are representations of the universal
covering group \(\Spin(n)\) of \(\SO(n)\), the numerators also become functions
of spinor operators built out of the Dirac (Clifford) algebra generators
\(\gamma_\mu\) that satisfy the anticommutation relations \(\{\gamma_\mu,
\gamma_\nu\} = 2g_{\mu\nu}\) where \(g\) is the metric tensor.  Efficient
algorithms are known for simplifying \(\gamma\) matrix expressions
\cite{kennedy81a, kennedy82c, kennedy82d} and have been implemented in the
REDUCE package CVIT \cite{ilyin:1989a}; these algorithms are based on reducing
the ``spin-network'' or ``birdtrack'' diagrams in terms of sums of irreducible
representations of \(\Spin(n)\) and their \(3j\) and \(6j\) coefficients.  We
are also currently investigating the generalization of these methods to handle
arbitrary representations of \(\SO(n)\), which might lead to much more
efficient algorithms for symbolic tensor manipulation than exist at present.

\begin{figure*}[ht]
  \begin{verbatim}
  diag2 := diagram(
    seq(vertex(cat('V',i),'ThreeScalarVertex'), i = 4..7),
    externalline(E4,V4,p3,'Scalar'), externalline(E5,V6,-p3,'Scalar'),
    line(I4,V4,V5,'Scalar'), line(I5,V4,V7,'Scalar'), line(I6,V5,V7,'Scalar'),
    line(I7,V5,V6,'Scalar'), line(I8,V6,V7,'Scalar'))$
  g2 := analyzeDiagram(diag2)$
  analyzeDiagram:   "2 loop diagram"
  rg2 := R(g2);
  R:   "Applying R to IPI graph Graph(EXT(E4,E5),INT(I4,I5,I6,I7,I8))"
  R:   "Applying R to IPI graph Graph(EXT(E4,I7,I8),INT(I4,I5,I6))"
  T:   "Taylor subtracting 1 loop graph Graph(EXT(E4,I7,I8),INT(I4,I5,I6)) of degree 0"
  R:   "Applying R to IPI graph Graph(EXT(E4,E5,I6),INT(I4,I5,I7,I8))"
  T:   "Taylor subtracting 1 loop graph Graph(EXT(E4,E5,I6),INT(I4,I5,I7,I8)) of degree -2"
  R:   "Applying R to IPI graph Graph(EXT(E5,I4,I5),INT(I6,I7,I8))"
  T:   "Taylor subtracting 1 loop graph Graph(EXT(E5,I4,I5),INT(I6,I7,I8)) of degree 0"
  T:   "Taylor subtracting 2 loop graph Graph(EXT(E4,E5),INT(I4,I5,I6,I7,I8)) of degree 2"
  T:   "Taylor subtracting 2 loop graph -Graph(EXT(E4,E5),INT(I4,I5,I6,I7,I8)) of degree 2"
  T:   "Taylor subtracting 2 loop graph -Graph(EXT(E4,E5),INT(I4,I5,I6,I7,I8)) of degree 2"
  \end{verbatim}
  \caption{Example of the use of our Maple program for the two-loop two-point
    function \(\diagchar{'00}\) in \(\phi^3\) theory.}
  \label{fig:ExampleMaple}
\end{figure*}

\break
\section{Infrared Cancellations}
The dedicated reader might have observed that we do not live in Euclidean
space, but in Minkowski space.\footnote{At least ``if we keep our feet on the
ground and ignore gravity'' (E.~Mottola).} Fortunately this has absolutely no
effect on the algebraic manipulations we have described: we end up with a
renormalized integrand depending on a set of \(\SO(4,\C)\) invariants
constructed out of the external momenta, such as \(p^2\), \(p\cdot q\),
\(\epsilon_{\mu\nu\rho\sigma} p^\mu q^\nu r^\rho s^\sigma\) and the difference
between the real forms \(\SO(4,\R)\) and \(\SO(3,1,\R)\) is just that \(p^2\)
can be negative, for example.

While the cancellation of UV divergences works just as well in Minkowski space
as it does in Euclidean space, new IR divergences can occur.  Previously we
evaded all IR divergences by keeping the mass \(m\) of our particles non-zero,
in the real world we need to handle the fact that some real particles are
massless (such as the photon\footnote{In fact probably only the photon, as
neutrinos seem to have masses, albeit very small ones.  Gluons, the gauge
bosons for QCD, are also massless; but as QCD is a strongly interacting
confined theory they do not exist as physical ``on-shell'' particles.}) and
also that there are a variety of other IR singularities possible because
\(k^2=0 \not\implies k=0\) in Minkowski space.

There are two different situations to consider:
\begin{itemize}
\item There are true IR divergences that make the theory meaningless.  For
  example, massless scalar field theory does not exist because it is IR sick in
  this manner.
\item There are IR singularities that occur at special values of the external
  momenta.  The emission of a zero-energy photon in QED exemplifies this
  situation.
\end{itemize}
We shall reject theories in the first case, whereas in the second we evade the
problem by insisting that on physical grounds one must average the
cross-section for emission of soft photons (Brehmsstrahlung) over some
experimental resolution for the detection of the photon energy.\footnote{In
mathematical terms we say that the Green's functions in Minkowski space are
(tempered) distributions rather than functions.}

The crucial observation is that while the Green's functions may have IR
divergences the cross-sections smeared over experimental resolutions are always
finite.  At the diagrammatic level what happens is that there are cancellations
between loop integrals (or Feynman parameter integrals) and phase space
integrals, so our intention is to extend our program to treat loop and phases
space integrals consistently.  What we must do is to ensure that all
divergences cancel locally in the integrand, just as we do for Green's
functions by our use of the \(R\) operation.  To this end we intend to generate
the integrands for cross sections by representing them as cut bubble diagrams,
where the cut propagators correspond to on-shell particles convoluted with the
experimental resolution functions.

\section{Computational Issues}
Our prototype program is written in {\tt Maple} and generates {\tt C} code: an
example of its use is illustrated in Figures~\ref{fig:ExampleMaple}
and~\ref{fig:ExampleC}.  This is unusual in that most symbolic computations of
Feynman diagrams make use of domain-specific languages such as {\tt FORM}
\cite{strubbe:1974a, vermaseren:1999a, vermaseren:2000a}, which are relatively
unsophisticated but can execute their limited repertoire of operations on very
large expressions with great efficiency.

\begin{figure}[ht]
  \begin{verbatim}
#include <math.h>
float term2(float x4,float x5,float x6,
	    float x7,float x8)
{
  float SMadjr1c1;
  float SMadjr1c2;
  float SMadjr2c2;
  float SmuSq;
  float t11;
  float t13;
  float t20;
  float t21;
  float t24;
  float t25;
  extern float m;
  extern float p3p3;
  {
    SMadjr1c1 = x4+x5+x6+x7+x8;
    SMadjr1c2 = -x4-x5;
    SMadjr2c2 = -SMadjr1c2;
    t11 = m*m;
    t13 = x4*x4;
    SmuSq = p3p3*x4+t11*SMadjr1c1-(2.0*
      SMadjr2c2*t13-(-SMadjr2c2*x4-x4*
      SMadjr1c1)*x4)*p3p3;
    t20 = 0.3141593E1*0.3141593E1;
    t21 = t20*t20;
    t24 = logf(SmuSq);
    t25 = sqrtf(x4*x6+x7*x4+x8*x4+x5*x6+x7*x5+
      x8*x5);
    return(-t21*t20*SmuSq*t24/t25);
  }
}
  \end{verbatim}
  \vskip-5ex
  \caption{Part of the {\tt C} code generated by the example of
    Figure~\ref{fig:ExampleMaple}.}
  \label{fig:ExampleC}
\end{figure}

\subsection{Code Generation}
We could try to evaluate the final numerical integrals within {\tt Maple}, but
other than for debugging purposes this is too inefficient even using the {\tt
NAG} integration routines packaged within {\tt Maple}.  We therefore need to
generate efficient numerical code to evaluate the integrand for use within our
special-purpose Monte Carlo integration programs.  It is not too important
which language is used, our example is in {\tt C} but it would be trivial to
generate {\tt Fortran} if that was preferred.


\subsection{Memory Management and Laziness}
The computational model that {\tt FORM} uses is to stream a sum of many terms
from disk to disk, either applying pat\-tern-based transformations or sorting
and collecting terms as the data passes through the processor.  This model is
justified by the huge size of the expressions that are generated in realistic
Feynman diagram computations, unlike the toy example of
Figure~\ref{fig:ExampleMaple}.  While our {\tt Maple} program works well on
moderate size problems it dies a horrible death when the entire problem no
longer fits in memory.

We suggest that the huge intermediate expressions generated are not really an
intrinsic property of the problem, but are an artefact of the way we think
about and implement the programs.  To be specific, our program first applies
the \(R\) operation generating a sum of Taylor subtraction terms as it recurses
through the Henge decomposition; it then runs through all these terms
converting them into Feynman parameter integrals; it simplifies the tensor
structures in the numerators; and it then collects all the terms with similar
quadratic forms in their denominator before finally generating {\tt C} code.
This multi-pass approach is not necessary, it is just convenient for thinking
about and debugging the program, as it separates the computation into logically
independent steps.

We could avoid generating huge intermediate expressions by generating the terms
lazily, and this could be done declaratively without altering the logic of the
program if {\tt Maple} implemented a lazy {\tt map} primitive, or a {\tt yield}
statement for use within loops.  We would then generate each term and
immediately apply subsequent operations on it until we reach the ``collection''
phase, in which terms could be collected into elements of a hashed array.

Eventually this approach will still run out of memory if this array has to live
in (virtual) memory, but this could be circumvented by writing the array
elements to disk when necessary.

\section{Conclusions}
Current perturbative calculations in QFT are unthinkable without the use of
computer algebra. Whether this requires domain-specific languages such as {\tt
FORM} or more careful memory choreography in general purpose systems such as
{\tt Maple} remains to be seen. All of the pieces of the calculation, including
renormalization, can be fully automated except for evaluation of the Feynman
parameter integrals where the lack of closed-form solutions forces us into the
arms of numerical integration.


\balancecolumns
\end{document}